\documentclass[12pt]{article}

\begin{document}

\title{Relativistic Quantum Finance }

\author{ Juan M. Romero \thanks{jromero@correo.cua.uam.mx} and Ilse B. Zubieta-Mart\'inez \\ 
\it Departamento de Matem\'aticas Aplicadas y Sistemas\\
\it Universidad Aut\'onoma Metropolitana-Cuajimalpa\\
\it M\'exico, D.F  01120, M\'exico\\}
%\\

\date{}

\pagestyle{plain}

\maketitle

\begin{abstract}
Employing the Klein-Gordon equation,  we propose a generalized Black-Scholes equation.
In addition, we found a limit  where this generalized equation is invariant under conformal transformations, in particular invariant under
scale transformations. In this limit, we show that  the stock prices distribution is given by a  Cauchy distribution, instead of a normal distribution.
\end{abstract}

\section{Introduction}

The Black-Scholes equation is one of the most useful in finance \cite{black,merton}. This equation can be obtained from different approaches,
 for example as a limit from Cox-Ross-Rubenstein model or from stochastic calculus.  
However,  Black-Scholes equation  is based in some ideal assumptions, for example that there are not arbitrage and that
the stock prices follow a normal distribution. Whereby,  in some cases the Black-Scholes equation   can not provide 
realistic predictions.  This fact has been noted by different authors.  Actually, before  the Black-Scholes equation  was proposed, Mandelbrot  noticed that some  stock  prices 
do not follow a normal distribution, but a Cauchy distribution \cite{mandelbrot}. In order to obtain a more realistic Black-Scholes equation, various authors  
haven been proposed different generalized Black-Scholes equations. 
 For instance using  a stochastic volatility \cite{heston}, multifractal volatility \cite{calvet}, jump processes \cite{tankov}, Levy's  distributions \cite{levy} and fractional differential equations \cite{kleinert1}. \\ 

Interesting, recently diverse mathematical techniques  of physics have been applied in finance successfully \cite{baaquie,bouchaud,voit}. 
For example,  quantum mechanics is a natural framework to study financial models with stochastic volatility or stochastic interest rate \cite{baaquie}. Furthermore, statistical mechanics can be used to study  financial risk \cite{bouchaud} and fluid theory can be employed to study foreign exchange markets \cite{voit}. Moreover, statistical arbitrage
can be studied with  relativistic statistical mechanics \cite{wissner}. Additionally  some financial crashes can be seen as a  phase transition \cite{crashes,stanley}, 
it is worth mentioning that 
when a system is in a phase transition some of  its  quantities are invariant under scale transformations \cite{cmatter}. In the relation  between physics and finance,  
a remarkably result is that the  Black-Scholes equation can be mapped  to the free Schr\"odinger equation \cite{baaquie}. Then, the Black-Scholes formula can be obtained using quantum mechanics.  It is worth mentioning that in this mapping the particle mass $m$ is related  with the volatility $\sigma$ in the way  
$m\to  1/\sigma^{2}.$ Thus,  an stock price with high volatility is identified with a light weight particle, while  a massive particle is identified with a stock price with small volatility. 
Now, it is well know that when $\sigma$ is too large  the Black-Scholes equation does not make sense. While, when  $m$ is very small the Schr\"odinger equation does not make sense too. In fact, in this last case the quantum mechanics is  changed by the relativistic quantum mechanics and  the Schr\"odinger is  changed by  the Klein-Gordon equation. In the Klein-Gordon equation a new parameter is introduced, the speed of light, $\tilde c.$ When  $\tilde c\to \infty$ we obtain  the Sch\"odinger equation.  \\

 In this paper, in order  to obtain a generalized Black-Scholes equation,
 we   relate   the Klein-Gordon equation with  a new generalized Black-Scholes equation. 
 We show that there is  a limit  where the generalized Black-Scholes equation is invariant under conformal transformations, in particular invariant under
scale transformations. In this limit, we show that  the stock prices distribution is given by a  Cauchy distribution, instead of a normal distribution.  Due that scale invariance is a characteristic  of phase transitions \cite{cmatter}, we can think that when the modified Black-Scholes equation is invariant under scale transformation   the system  is  near from a phase transition. \\

This paper is organized as follow:  In  Section 2, it is shown the mapping between the free Schr\"odinger equation and the  Black-Scholes equation; in Section 3, 
it is proposed a relativistic Black-Scholes equation; in Section 4, the conformal symmetry is studied; in Section 5, the Cauchy distribution is obtained. Finally, in Section 6 a summary is given.

\section{Black-Scholes equation and free Schr\"odinger equation }

The  Black-Scholes equation is given by 
\begin{eqnarray}
\frac{\partial C(S,t) }{\partial t}= -\frac{\sigma ^{2}}{2}S^{2} \frac{\partial^{2} C(S,t) }{\partial S^{2}}-r S \frac{\partial^{2} C(S,t) }{\partial S}
+rC(S,t), \label{bse0}
\end{eqnarray}
where 
$\sigma$ is the volatility, $S$ is the stock price, $r$ is the annualized  risk-free inters rate  and $C$ is the option   price. While the free Schr\"odinger equation is 
\begin{eqnarray}
i\hbar \frac{\partial \psi }{\partial \tilde t}= -\frac{\hbar ^{2}}{2m}\frac{\partial^{2} \psi }{\partial x^{2}}, \label{eq:sch}
\end{eqnarray}
here $m$ is the particle mass, $\hbar$ is the Planck constant and $\psi$ is the wave function.\\

Remarkably,  using the mapping
\begin{eqnarray}
\tilde t&=&it, \quad \hbar=1, \quad m=\frac{1}{\sigma^{2}}, \qquad x=\ln S, \\
\psi(x,t)&=& e^{-\left(\frac{1}{\sigma^{2}}\left(\frac{\sigma^{2}}{2}-r \right)x +\frac{1}{2\sigma^{2}} \left(\frac{\sigma^{2}}{2}+r \right)^{2} t\right)}C(x,t)
\label{eq:map}
\end{eqnarray}
the free Schr\"odinger equation (\ref{eq:sch}) becomes  the Black-Scholes equation (\ref{bse0}).\\

\section{Relativistic quantum finance}

It is well known that when $m\to 0$ the Schr\"odinger equation does note make sense.  While, if $\sigma^{2}$ is not small the BS equation does not  make sense  too. 
In physics the case $m\to 0$ is not studied in the  usual  quantum mechanics, but in the relativistic  quantum mechanics. In this last theory, the Schr\"odinger
equation is changed by the Klein-Gordon equation
\begin{eqnarray}
-\frac{\hbar^{2}}{\tilde c^{2} } \frac{\partial^{2} \psi(x,t) }{\partial \tilde t^{2}}+\frac{\partial^{2} \psi(x,t) }{\partial x^{2}}-m^{2}\tilde c^{2}\psi(x,t)=0,\label{eq:kge}
\end{eqnarray}
 where $\tilde c$ is the light speed. Notice  that when $m\to 0,$ the Klein-Gordon equation makes sense. Furthermore, it can be shown that
 when $ \tilde c\to \infty$ the Klein-Gordon equation becomes the Schr\"odinger equation.\\
 
 Now, using the mapping 
\begin{eqnarray}
 \tilde t&=&it, \quad \hbar=1, \quad m=\frac{1}{\sigma^{2}}, \qquad x=\ln S, \\ 
 \tilde c^{2}&=&q, \qquad \psi(x,t)= e^{-\left[\frac{1}{\sigma^{2}}\left(\frac{\sigma^{2}}{2}-r \right)x +\left( \frac{1}{2\sigma^{2}} \left(\frac{\sigma^{2}}{2}+r \right)^{2} -\frac{q}{\sigma^{2}} \right)  t\right]}C(x,t)
\label{map}
\end{eqnarray}
 the Klein-Gordon equation becomes
\begin{eqnarray}
\frac{1}{q} \frac{\partial^{2} C(S,t) }{\partial t^{2}}+\left( \frac{2}{\sigma^{2} }-\frac{1}{q\sigma^{2}}\left( \frac{\sigma^{2}}{2}+r\right)^{2} \right)  \frac{\partial C(S,t) }{\partial t}+S^{2} \frac{\partial^{2} C(S,t) }{\partial S^{2}}\nonumber \\+\frac{2r}{\sigma^{2}} S \frac{\partial C(S,t) }{\partial S}
+
\left[  \frac{1}{4q \sigma^{4} }\left( \frac{\sigma^{2}}{2}+r\right)^{4}- \frac{2r}{\sigma^{2}}  \right]C(S,t)=0.\label{mbs}
\qquad
\end{eqnarray}
This  equation can be written as 
 \begin{eqnarray}
& & \frac{\sigma^{2} }{2q } \frac{\partial^{2} C(S,t) }{\partial t^{2}}+\left( 1-\frac{1}{2q}\left( \frac{\sigma^{2}}{2}+r\right)^{2} \right) \frac{\partial C(S,t) }{\partial t}\nonumber\\
&=& -\frac{\sigma^{2}}{2} S^{2} \frac{\partial^{2} C(S,t) }{\partial S^{2}}-r S \frac{\partial C(S,t) }{\partial S}
+rC(S,t).\label{eq:mbs1}
\end{eqnarray}
Furthermore, if we take the limit  $ q\to  \infty,$  in this last  equation  we arrive to the Black-Scholes equation (\ref{bse0}). Then, the equation (\ref{mbs}) is a generalized  Black-Scholes equation, which has the new parameter $q.$ \\

\section{Conformal symmetry}

The usual Black-Scholes equation is invariant under the Schr\"odinger group \cite{nos}.\\

In orden to understand  the symmetries for the equation (\ref{mbs}) we take  the coordinates  
\begin{eqnarray}
z=\ln S+ i\sqrt q t.
\end{eqnarray}
Using this coordinates,  the modified Black-Scholes equation (\ref{mbs}) can be written as  
\begin{eqnarray}
4\frac{\partial^{2}C }{\partial z\partial \bar z}+2\left( \bar A\frac{\partial C}{\partial z}+   A\frac{\partial C}{\partial \bar z}\right)+ \left(  A\bar A- \frac{q}{\sigma^{4}}\right)C=0,
\label{eq:cmbs}
\end{eqnarray}
where 
\begin{eqnarray}
A=-\frac{1}{\sigma^{2}} \left[ \left(\frac{\sigma^{2}} {2} -r\right)-\frac{i}{2 \sqrt{q}} \left( \left( \frac{\sigma^{2}} {2} +r \right)^{2} -2q\right) \right].
\end{eqnarray}
We can see that  the equation (\ref{eq:cmbs}) is invariant under the following  transformations
\begin{eqnarray}
z^{\prime}&=&e^{i \alpha} z,\qquad \alpha={\rm constant},\\ 
C^{\prime}\left(z,\bar z^{\prime}\right)&=& e^{\frac{1}{2} \left[ Az\left(1-e^{i\alpha} \right) +\bar A \bar z\left(1-e^{-i\alpha} \right) \right] }C(z,\bar z).
\end{eqnarray}

We have a special case, in fact when 
\begin{eqnarray}
  \frac{q}{\sigma^{4}}<< \bar AA,
\end{eqnarray}  
that is 
\begin{eqnarray}
 0 << \frac{1}{4q }\left( \frac{\sigma^{2}}{2}+r\right)^{4}+2r \sigma^{2}, \label{eq:ft}
\end{eqnarray} 
 the equation (\ref{eq:cmbs}) becomes 
\begin{eqnarray}
4\frac{\partial^{2}C }{\partial z\partial \bar z}+2\left( \bar A\frac{\partial C}{\partial z}+   A\frac{\partial C}{\partial \bar z}\right)+  A\bar A C=0.
\label{eq:cbs}
\end{eqnarray}
This last equation is invariant under the conformal symmetry 
\begin{eqnarray}
z^{\prime}&=&z^{\prime}(z), \\
C^{\prime}\left(z^{\prime} ,\bar z^{\prime}\right)&=& e^{\frac{1}{2} \left[ A\left(z-z^{\prime} \right) +\bar A\left( \bar z-\bar z^{\prime} \right) \right]}C(z,\bar z). \label{tp}
\end{eqnarray}
 where 
\begin{eqnarray}
 \frac{\partial z^{\prime} (z) }{\partial \bar z}=0, \label{con}
 \end{eqnarray}
namely $z^{\prime} (z)$ is an analytic function. Then, every analytic  function  provides a symmetry for the equation (\ref{eq:cbs}). \\

Notice that the equation (\ref{con}) implies  that at first order any infinitesimal transformation can  be expressed as  
\begin{eqnarray}
 z^{\prime}= z+\epsilon(z), \qquad \epsilon(z)=\sum_{-\infty}^{\infty} \epsilon_{n} z^{n+1}, \qquad \epsilon_{n}={\rm constant}. \label{ict}
\end{eqnarray}
Now, using the equation (\ref{tp}), we find  
\begin{eqnarray}
 C^{\prime}\left(z^{\prime}, \bar z^{\prime}\right)\approx  C\left(z^{\prime},\bar z^{\prime}\right)-\left[ \epsilon\left(z^{\prime} \right)\left( \frac{A}{2}+\frac{\partial }{\partial z^{\prime} }\right) 
 + \bar \epsilon\left(\bar z^{\prime} \right)\left( \frac{\bar A}{2}+\frac{\partial }{\partial \bar z^{\prime} }\right) \right] C\left(z^{\prime},\bar z^{\prime}\right),
\end{eqnarray}
that is 
\begin{eqnarray}
 \delta C\left(z^{\prime}, \bar z^{\prime}\right)&=& C^{\prime}\left(z^{\prime}, \bar z^{\prime}\right)- C\left(z^{\prime},\bar z^{\prime}\right)\nonumber\\
 &\approx &-\left[ \epsilon\left(z^{\prime} \right)\left( \frac{A}{2}+\frac{\partial }{\partial z^{\prime} }\right) 
 + \bar \epsilon\left(\bar z^{\prime} \right)\left( \frac{\bar A}{2}+\frac{\partial }{\partial \bar z^{\prime} }\right) \right] C\left(z,\bar z\right).
\end{eqnarray}
Furthermore, using the equation (\ref{ict}) the option price transforms as
\begin{eqnarray}
 \delta C\left(z, \bar z\right)&=&\sum_{-\infty}^{\infty}\left(  \epsilon_{n} l_{n} +    \bar \epsilon_{n} \bar l_{n} \right) C\left(z^{\prime},\bar z^{\prime}\right),
\end{eqnarray}
where 
\begin{eqnarray}
l_{n}=-z^{n}\left( \frac{A}{2}+ \frac{\partial }{\partial z}\right), \qquad   \bar l_{n}=-\bar z^{n}\left( \frac{\bar A}{2}+ \frac{\partial }{\partial \bar z}\right).
\end{eqnarray}
These operator satisfy the Witt algebra 
\begin{eqnarray}
\left [ l_{n}, l_{k}\right ] &=&(n-k)l_{n+k},\nonumber \\
\left [ \bar l_{n}, \bar l_{k}\right] &=&(n-k)\bar l_{n+k},\nonumber \\
\left[ \bar l_{n},  l_{k}\right ] &=&0.
\end{eqnarray}

Notably, the usual Black-Scholes
equation has only a finite number of symmetries, but the equation (\ref{eq:cbs})  has an infinity number of symmetries.\\

In particular, if $\lambda$  is a real number, the equation (\ref{eq:cbs}) is invariant under the scale transformations 
\begin{eqnarray}
z^{\prime}&=&\lambda z, \\
C^{\prime}\left(z^{\prime} ,\bar z^{\prime} \right)&=& e^{\frac{1}{2} (1-\lambda ) \left(Az +\bar A\bar z\right)} C(z,\bar z),
\end{eqnarray}
 which can be written as 
\begin{eqnarray}
S^{\prime}&=&S^{\lambda}, \\
t^{\prime}&=&\lambda t ,\\
C^{\prime}\left(S^{\prime} , t^{\prime} \right)&=& C^{\prime}\left(S^{\lambda}, \lambda t\right)= S^{\frac{(\lambda -1)}{\sigma^{2}} \left( \frac{\sigma^{2} }{2}-r\right) }e^{\frac{(\lambda-1 ) }{2\sigma^{2}} \left( \left( \frac{\sigma^{2} }{2}+r\right)^{2}-2q\right) t } C(S,t).
\end{eqnarray}
In general, the equation (\ref{mbs}) is not invariant under scale transformations, however in the limit (\ref{eq:ft}) this symmetry is obtained. This phenomenon   is well known in phase transitions, in fact the scale invariance is a characteristic  of phase transitions \cite{cmatter}. Then, we can think that when the equation (\ref{eq:ft}) is satisfied,  the system  is  near from a phase transition.  Remarkably, when a system is near from a phase transition there are fluctuations of all scales. This phenomenon happened  in some  financial crashes \cite{crashes,stanley}.

\section{ Cauchy distribution}

Using the coordinates $x, t$ and the function $\psi$ defined in  (\ref{map}), the equation (\ref{eq:cbs}) can be written as  
\begin{eqnarray}
\frac{1}{q } \frac{\partial^{2} \psi(x,t) }{\partial  t^{2}}+\frac{\partial^{2} \psi(x,t) }{\partial x^{2}}=0. \label{eq:lap}
\end{eqnarray}
Now, imposing the initial condition
\begin{eqnarray}
\psi(x,0)=f(x),
\end{eqnarray}
the solution for the  equation (\ref{eq:lap})  is given by
\begin{eqnarray}
\psi(x,t)=\int_{-\infty }^{\infty} d\zeta  G(x-\zeta,t)f(\zeta),
\end{eqnarray}
where 
\begin{eqnarray}
G(x-\zeta,t)= \frac{1}{\pi} \frac{  \sqrt{q}t  }{(x-\zeta)^{2}+qt^{2}  } .
\end{eqnarray}
Notice that this last function is the Cauchy distribution.   Using the mapping (\ref{eq:map}), we obtain the  option price 
\begin{eqnarray}
C(x,t)=\int_{-\infty }^{\infty} d\zeta  K(x-\zeta,t)C(\zeta,0),
\end{eqnarray}
here 
\begin{eqnarray}
 K(x-\zeta,t)=
  \frac{ e^{\frac{1}{\sigma^{2} }\left[  \left(\frac{1}{2} \left( \frac{\sigma^{2}}{2}+r\right)^{2}-q \right) t +\left( \frac{\sigma^{2}}{2} -r\right) (x-\zeta )  \right]}  }{\pi}   \frac{  \sqrt{q}t  }{(x-\zeta)^{2}+qt^{2}  } .
\end{eqnarray}
The Cauchy  distribution was  first proposed by B. Mandelbrot as a distribution for stock price \cite{mandelbrot}. 

\section{Summary}
Employing the Klein-Gordon equation,  we proposed a generalized Black-Scholes equation. 
We found a limit  where the generalized equation is invariant under conformal transformations, in particular invariant under
scale transformations. In this limit, we shown that  the stock prices distribution is given by a  Cauchy distribution, instead of a normal distribution.

\end{document}